\documentclass{aa}
\usepackage{graphicx}
\usepackage{txfonts}
\def\lsim{\mathrel{\rlap{\lower2.5pt\hbox{\hskip1pt$\sim$}}
    \raise1pt\hbox{$<$}}}         
\def\gsim{\mathrel{\rlap{\lower2.5pt\hbox{\hskip1pt$\sim$}}
    \raise1pt\hbox{$>$}}}         
\begin{document}
\title{Stability of the mixed phase in hybrid stars}

\author{Omar Benhar \inst{1,2} \and Roberto Rubino\inst{2}}

\institute{
INFN, Sezione di Roma. Piazzale Aldo Moro, 2, I-00185 Roma, Italy
\and
Dipartimento di Fisica, Universit\`a ``La Sapienza". Piazzale Aldo Moro, 
2, I-00185 Roma, Italy
}

\date{Received ; accepted }

\abstract{

The transition from hadronic matter to quark matter in the core of neutron 
stars is likely to be associated with the appearance of a mixed phase, leading to 
a smooth variation of the star density profile. We discuss the results of 
a systematic study of the properties of the mixed pase upon Coulomb and 
surface effects. A state of the art nonrelativistic equation of state of 
nuclear matter has been used for the low density phase, while 
quark matter has been described within the MIT bag model, including the 
effect of perturbative one-gluon exchange interactions. The implications for 
neutron star structure are discussed.

   
}

\maketitle

\section{Introduction}
\label{intro}

The possible occurrence of a core of deconfined quark matter in the interior of 
neutron stars has been investigated by a number of authors over the past three decades
(for a recent review see, e.g., Weber \cite{weber}).
Due to the complexity of the underlying dynamics, theoretical 
approaches to the study of hybrid stars largely rely on models 
to describe the equation of state (EOS) of strongly
interacting matter, in both the hadronic and quark sector, as well as on 
a set of assumptions on the nature of the phase transition.

Most calculations have been carried out using nuclear EOS obtained from
either nonrelativistic nuclear many body theory (NMBT) or relativistic 
mean field theory (RMFT), while the deconfined phase is usually described 
within the MIT bag model (Chodos et al \cite{bagmodel}).

In their pioneering work Baym and Chin (\cite{baym}) employed the familiar 
Maxwell double tangent construction (see, e.g., Huang \cite{huang}), which
amounts to assume that the transition occur at constant pressure. Within this
picture, charge neutral nuclear matter at energy density 
$\epsilon_{NM}$ cohexists with charge neutral quark matter at energy density 
$\epsilon_{QM}$, the two phases being separated by a sharp interface.

In the early 90s Glendenning (Glendenning \cite{glend1,glend2}) first pointed out 
that the requirement that the two phases be individually charge neutral is in fact 
too restrictive. In a more general scenario charged nuclear and quark matter 
may share a common lepton background, thus giving rise to a mixed phase extending 
in space over a sizable fraction of the star. 

The appearance of a mixed phase strongly affects the macroscopic properties
of the star. A transition at constant pressure necessarily leads to
the appearance of a discontinuity in the density profile, i.e. to a star
consisting of a inner core of quark matter at energy density $\epsilon_{QM}$ 
surrounded by nuclear matter at energy density $\epsilon_{NM}$.
On the other hand, the mixed phase allows for a smooth variation              
of the energy density, leading in turn to a smooth variation of the star 
density profile.      

Whether the transition proceeds at constant pressure or according to Glendenning's 
picture depends upon i) the value of the Debye screening length, driving charge 
separation, and ii) the amount of electrostatic and surface energy needed for the 
formation of the structures of quark and nuclear matter in the mixed phase. If 
screening is too strong to allow for a uniform lepton density, or if 
the energy loss due to Coulomb and surface effects exceeds the gain in
bulk energy, the standard scenario predicted by Maxwell construction
turns out to be favorable.

The results of a detailed study carried 
out by Heiselberg et al (\cite{HPS93}), suggest that the mixed phase
is energetically favored over a sizable density range if the surface tension 
is less that $\sim 70$ MeV/fm$^2$. In their calculations these authors
adopted the MIT bag model (Chodos et al \cite{bagmodel}) for the quark matter 
EOS, while nuclear matter was described using a somewhat oversimplified model, 
in which the energy-density included a quadratic compressional term 
and a symmetry term taken from 
a previous work of Lattimer et al (\cite{compress}).

Over the past decade, the availability of new nucleon-nucleon potentials, resulting 
from accurate fits to nucleon-nucleon scattering data, and the improvement of the 
computational schemes made it possible to obtain a new generation of EOS within 
the framework of NMBT (Akmal \& Pandharipande \cite{apr}, Akmal et al
 \cite{apr2}). In view of the fact that NMBT is a parameter free approach, 
whose dynamics is strongly cosntrained by nuclear data, and has been shown to possess a
highly remarkable predictive power in theoretical studies of few nucleon systems
(Pieper \& Wiringa \cite{pw}), it provides a natural candidate to describe neutron 
star matter in the nuclear phase. 

In this paper we extend the work of Heiselberg et al (\cite{HPS93}) 
carrying out a systematic study of the stabilty of the mixed phase. We adopt the 
state of the art EOS of nuclear matter, obtained by 
Akmal et al (\cite{apr2}) within NMBT, and analyze the 
dependence of the results on i) the parameters entering the MIT bag model EOS, 
employed to describe quark matter, and ii) the value
of the surface tension, driving both Coulomb and surface effects.

In Section \ref{EOS} we summarize the main features of the model EOS of 
both nuclear and quark matter, while the implementation of 
Gibbs conditions in the case of two chemical potentials, leading 
to the appearance of the mixed phase, is discussed in Section 
\ref{phase_transition}. Our main results are presented 
in Section \ref{surface}, devoted to the role of Coulomb and surface energy.
The implication of the appearance of the mixed phase for neutron star structure
are outlined in Section \ref{star}.
Finally, the conclusion of our work are stated in Section \ref{conclusions}.

\section{Models of neutron star matter EOS}
\label{EOS}

In this Section we summarize the main features of the 
EOS employed in our work, focusing on the region of 
nuclear and supranuclear density ($n_B \gsim 0.1$ fm$^{-3}$). For
the lower density region, corresponding to the 
outer and inner crust of the star, we have used the EOS of Baym et al
 (\cite{BPS}) and Pethick et al (\cite{PRL}), 
respectively. It has to be pointed out, however, that our results 
are largely unaffected by the details of the EOS at subnuclear density, as 
the fraction of star mass in the crust is only about 2 \%.

The theoretical descriptions of nuclear and quark matter are both 
based on the standard assumptions that the system be at zero temperature 
and transparent to neutrinos produced in weak interaction processes 
(see, e.g., Shapiro \& Teukolski \cite{shapiro}).

\subsection{Nuclear matter}
\label{EOS_nm}

Within NMBT, nuclear matter is 
viewed as a collection of pointlike protons and neutrons, whose dynamics
is described by the hamiltonian
\begin{equation}
H = \sum_{i} \frac{ {\bf p}_i^2 }{ 2 M } + \sum_{j>i} v_{ij}
+ \sum_{k>j>i} V_{ijk}  \ , 
\label{nucl:ham}
\end{equation}
$M$ and ${\bf p}_i$ being the nucleon mass and the momentum of the i-th 
particle, respectively. The nucleon-nucleon (NN) potential $v_{ij}$
reduces to the well known Yukawa one-pion exchange potential at large 
distances, while 
its short and intermediate range behavior is determined by fitting
the available experimental data on the two-nucleon system (deuteron
properties and $\sim$ 4000 NN scattering phase shitfs) 
(Wiringa et al \cite{av18}). 
The three-nucleon potential $V_{ijk}$, whose inclusion is needed 
to reproduce the observed binding energies of the three-nucleon system, 
consists of the Fujita-Miyazawa two-pion exchange potential supplemented 
by a purely phenomenological repulsive interaction 
( Pudliner et al \cite{uix}).

The many-body Schr\"odinger equation associated with the hamiltonian
of Eq.(\ref{nucl:ham}) can be solved exactly, using stochastic methods,
for nuclei of mass number A $\le 10$. The resulting energies of the ground and
low-lying excited states are in excellent agreement with experimental
data (Wiringa \& Pieper \cite{pw}). Exploiting translational invariance, 
accurate calculations can also be carried out for uniform nuclear matter 
(Wiringa et al \cite{wff} , Akmal \& Pandharipande \cite{apr}).

Akmal \& Pandharipande (\cite{apr}) have used cluster expansions and chain 
summation techiniques to obtain the energy per particle of 
both pure neutron matter (PNM) and symmetric nuclear matter (SNM).
In their approach the Argonne $v_{18}$ potential of Wiringa et al (\cite{av18}) 
is modified to take into account the fact that
NN potentials fitted to scattering data describe interactions between nucleons 
in their center of mass frame, in which the total momentum ${\bf P}$ vanishes.
Within the approach of Akmal \& Pandharipande relativistic corrections arising 
from the boost to a frame in which ${\bf P} \neq 0$, are included up to 
order ${\bf P}^2/M^2$.

Interpolating between the PNM and SNM results of
Akmal \& Pandharipande (\cite{apr}), Akmal et al (\cite{apr2}) have 
determined the energy of matter with arbitrary 
ptoton fraction, needed to obtain the EOS of $\beta$-stable matter, consisting of 
neutrons, protons electrons and muons. Their calculations span
 a range of baryon number density $n_B$ extending up to 
$\sim$ 8 $n_0$, $n_0 = 0.16$ fm$^{-3}$ being the empirical saturation density 
of symmetric nuclear matter.

At any given value of $n_B$ proton and lepton densities are determined
by the requirements of charge neutrality,
\begin{equation}
n_p = n_e + n_\mu \ ,
\end{equation}
and $\beta$-stability,
\begin{equation}
\mu_n = \mu_p + \mu_e \ ,
\end{equation}
\begin{equation}
\mu_e = \mu_\mu \ .
\end{equation}
In the above equations $n_i$ and $\mu_i$ denote the
number density and chemical potential of the particle of type $i$ $(i=n,p,e,\mu)$, 
respectively.

The EOS $P = P(\epsilon)$, where $P$ and $\epsilon$ denote pressure and
energy density is obtained from the density dependence of the binding energy
per baryon, $E = E(n_B)$, through
\begin{equation}
P = - \frac{1}{n_B^2}\frac{\partial E}{\partial n_B}\ ,
\label{eos1}
\end{equation}
and
\begin{equation}
\epsilon = n_B ( E + M ) \ .
\label{eos2}
\end{equation}

\begin{figure}[hbt]
\centering
\includegraphics[width=7.cm]{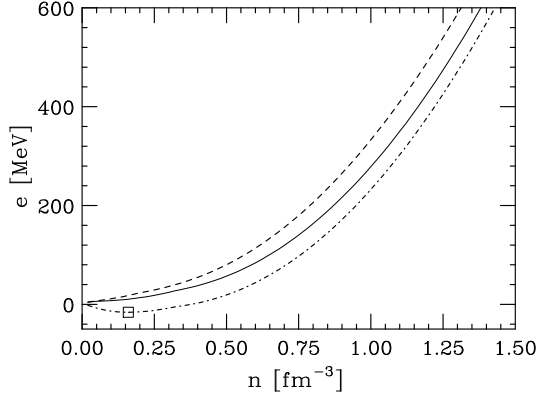}
\caption{
Energy per baryon of nuclear matter calculated by 
Akmal et al 
(\protect\cite{apr2}),plotted as a function of
baryon number density.
The dashed, dotdash and solid lines correspond to pure neutron matter, 
symmetrical nuclear metter and $\beta$-stable matter, respectively.
The box represents the equilibrium properties on symmetrical nuclear 
matter obtained from extrapolation of empirical data.
}
\label{aprenergy}
\end{figure}

The results discussed in the following sections have been obtained using
the EOS of Akmal {\it et al} (\cite{apr2}), hereafter referred
to as APR, to describe the nuclear phase of neutron star matter.
The corresponding energy per baryon of PNM, SNM and $\beta$-stable matter
is displayed in Fig. \ref{aprenergy} as a function of density.

\subsection{Quark matter}
\label{EOS_QM}

Due to the complexity of the fundamental theory of
strong interactions (Quantum Chromo-Dynamics, or QCD) 
a first principle description of the EOS of quark matter
at large density and low temperature is out of reach of the
existing computational approaches.
 To describe the quark matter phase, we have used the simple MIT bag model
(Chodos et al \cite{bagmodel}), 
in which the main features of QCD are implemented 
through the assumptions that: 
i) quarks occur in color neutral clusters confined to a finite region of space (the bag), 
whose volume is limited by the pressure of the QCD vacuum (the bag constant $B$),
and ii) residual interactions between quarks are weak, and can be treated 
in low order perturbation theory.

Within the MIT bag model the thermodynamic potential $\Omega$ can be written
as
\begin{equation}
\Omega = \Omega_{pert} + VB,
\end{equation}
where $\Omega_{pert}$ and $VB$, $V$ being the normalization volume, 
denote the perturbative and nonperturbative contribution, respectively,
and
\begin{equation}
\Omega_{pert} = V\ \sum_f \sum_n  \Omega^{(n)}_f\ .
\end{equation}
In the above equation, the index $f$ specifies the quark flavor, while 
$\Omega^{(n)}_f$ is the $n$-th order term of the perturbative 
expansion in powers of the strong coupling constant, $\alpha_s$. 

The EOS of quark matter can be obtained from the relations
linking pressure and energy density to $\Omega$:
\begin{equation}
P = - \frac{\Omega}{V} = 
- B - \sum_f \sum_n \Omega^{(n)}_f\ .
\label{def:P}
\end{equation}
\begin{equation}
\epsilon = 
\frac{\Omega}{V} - \sum_f \left( \frac{\partial \Omega}{\partial \mu_f} \right)
= -P + \sum_f \mu_f n_f\ ,
\label{def:eps}
\end{equation}
$n_f$ and $\mu_f$ being the density and chemical potential of the quarks of flavor $f$.

The lowest order perturbative contributions to the thermodynamic potential, 
corresponding to 
$\Omega^{(n)}_f$ with $n$ = 0 and 1, read (see, e.g., Tamagaki \& Tatsumi 
\cite {tamagaki})
\begin{eqnarray}
\nonumber
\Omega_0^f & = & 
-\frac{1}{\pi^2} \Bigg[ \frac{1}{4}\mu_f\sqrt{\mu_f^2-m_f^2}
\bigg( \mu_f^2 - \frac{5}{2} m_f^2 \bigg) \Bigg. \\
& & \ \ \ \ \ \ \ \ \ \ \ \ \ \ \ \ \ \ \ \ \ \ \ \ \ \ \ \ + \Bigg.
\frac{3}{8} m_f^4 \ln \frac{\mu_f-\sqrt{\mu_f^2-m_f^2}}{m_f}
 \Bigg]\ ,
\label{def:omega0}
\end{eqnarray}
\begin{eqnarray}
\nonumber
\Omega_1^f & = & \frac{2\alpha_s}{\pi^3}
\Bigg[ \frac{3}{4} \bigg( \ \mu_f\sqrt{\mu_f^2-m_f^2}
 -m_f^2 \ln \frac{\mu_f+\sqrt{\mu_f^2-m_f^2}}{m_f}\
\bigg)^2 \Bigg. \\
&  & \ \ \ \ \ \ \ \ \ \ \ \ \ \ \ \ \ \ \ \ \ \ \ \ \ \ \ \ \
- \frac{1}{2} \Bigg. (\mu_f^2-m_f^2)^2 \Bigg]\ .
\label{def:omega1}
\end{eqnarray}
The chemical potentials appearing in Eqs.(\ref{def:omega0}) 
and (\ref{def:omega1}) can be written
\begin{equation}
\mu_f = e_{F_f} + \delta\mu_f = \sqrt{ m_f^ 2 + p_{F_f}^2 } + \delta\mu_f, 
\end{equation}
where the first term is the Fermi energy of a gas of noninteracting 
quarks of mass $m_f$ at density $n_f = p_{F_f}^3/\pi^2$, whereas the
second term is a perturbative correction of order $\alpha_s$, whose
explicit expression is ( Baym \& Chin \cite{baymchin2})
\begin{equation}
\delta\mu_f = \frac{2\alpha_s}{3 \pi^2}  \left[ p_{F_f} - 
3 \frac{m_{F_f}^2}{e_{F_f}} 
\ln{ \left( \frac{ e_{F_f} + p_{F_f} }{m_f} \right) } \right]\ .
\end{equation}
In the case of massless quarks Eqs.(\ref{def:P}) and (\ref{def:eps}) 
reduce to the simple form
\begin{equation}
P = \frac{1}{4\pi^2} \left( 1 - \frac{2\alpha_s}{\pi} \right)
\sum_f \mu_f^4 - B
\end{equation}
\begin{equation}
\epsilon = \frac{3}{4\pi^2} \left( 1 - \frac{2\alpha_s}{\pi} \right)
\sum_f \mu_f^4 + B \ ,
\end{equation}
implying the relationship
\begin{equation}
P = \frac{\epsilon - 4B}{3}\ ,
\end{equation}
reminiscent of the EOS of noninteracting quarks.

For any baryon desity, quark densities are dictated by the requirements of 
baryon number conservation, charge neutrality and weak equilibrium. 
In the case of two flavors, in which only the light up and down quarks 
are present, we have
\begin{equation}
n_B = \frac{1}{3} (n_u + n_d)\ ,
\label{bnc:2}
\end{equation}
\begin{equation}
\frac{2}{3} n_u - \frac{1}{3}n_d - n_e = 0
\label{cn:2}
\end{equation}
\begin{equation}
\mu_d  = \mu_u + \mu_e\ ,
\label{we:2}
\end{equation}
where $n_e$ and $\mu_e$ denote the density and chemical potential of the
electrons produced through $d \rightarrow u + e^- + {\overline \nu}_e$.
Note that we have not taken into account the possible appearance of
muons, as in the relevant density region $\mu_e$ never exceeds the muon 
mass.

As the baryon density increases, the $d$-quark chemical potential 
reaches the value $\mu_d = m_s$, $m_s$ being the mass of the strange quark. 
The energy of quark matter can then be lowered turning d-quarks into 
$s$-quarks through $d + u \rightarrow u + s$. 
In presence of three flavors, Eqs.(\ref{bnc:2})-(\ref{we:2}) become
\begin{equation}
n_B = \frac{1}{3} (n_u + n_d + n_s)\ ,
\label{bnc:3}
\end{equation}
\begin{equation}
\frac{2}{3} n_u - \frac{1}{3}n_d - \frac{1}{3}n_s - n_e = 0
\label{cn:3}
\end{equation}
\begin{equation}
\mu_d  = \mu_s = \mu_u + \mu_e\ .
\label{we:3}
\end{equation}

Unfortunately, the parameters entering the bag model
EOS are only loosely constrained by phenomenology and their 
choice is somewhat arbitrary. 

As quarks are confined and not observables
as individual particles, their masses are not directly measurable
and must be inferred from hadron properties.
The Particle Data Group (Hagiwara {\it et al} \cite{PDG}) reports
masses of few MeV for up and down quarks and 60 to 170 MeV for the 
strange quark. We have set $m_u=m_d=0$ and $m_s=150$ MeV 
for the up, down and strange quark, respectively. In the density region 
relevant to our work heavier quarks do not play any role.

The strong coupling constant $\alpha_s$ can be obtained from the renormalization
group equation, yielding
\begin{equation}
\alpha_s = \frac{12 \pi}{ (33-2N_f) \ln{ ({\overline \mu^2}/\Lambda^2) } }\ ,
\end{equation}
where $N_f = 3$ is the number of active flavors, $\Lambda$ is the QCD scale 
parameter and ${\overline \mu}$ is an energy scale typical of the relevant 
density region (e.g. the average quark chemical potential). 
Using $\Lambda \sim 100 \div 200$ MeV (see, e.g., Ellis et al \cite{lambda}) 
and setting ${\overline \mu} = \mu_d \sim \mu_u$ at a typical baryon density 
$n_B \sim 4n_0$ one gets $\alpha_s \sim 0.4 \div 0.6$. The results discussed
in this paper have been obtained with $\alpha_s = 0.5$.

The values of the bag constant resulting from fits of the hadron spectrum 
range between $\sim 57$ MeV/fm$^3$, with $\Lambda = 220$ MeV, (De Grand et al 
 \cite{bval1}) and $\sim$ 350 MeV/fm$^3$ , with $\Lambda = 172$ MeV (Carlson, 
 et al \cite{bval2}). However, the requirement that the deconfinement 
transition do not occur at density $\sim n_0$ constrains $B$ to be larger than 
$\sim 120 - 150$ MeV/fm$^3$, and lattice results suggest a value of $\sim 210$ MeV/fm$^3$
(Satz \cite{bval3}). 
In order to gauge the dependence of the results upon the value of $B$, 
we have carried out our calculations setting $B = 120$ and $200$ MeV/fm$^3$.

\begin{figure}[hbt]
\centering
\includegraphics[width=7.cm]{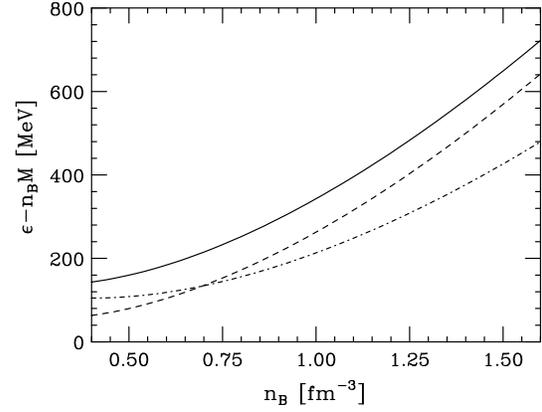}
\caption{Energy density of neutral quark matter in weak equilibrium as a
function of baryon number density. The solid and dashed lines have been obtained
setting $\alpha_s = 0.5$ and $B$ = 200 and 120 MeV/fm$^3$, respectively, while
the dashdot line corresponds to $\alpha_s = 0$ and $B = 200$ MeV/fm$^3$.
The quark masses are $m_u = m_d = 0$, $m_s = 150$ MeV.
}
\label{bagenergy}
\end{figure}

Fig. \ref{bagenergy} shows the energy density of neutral quark matter in weak 
equilibrium as a function of baryon density for different values of $B$
and $\alpha_s$. The solid and dashed lines have been obtained
setting $\alpha_s = 0.5$ and $B$ = 200 and 120 MeV/fm$^3$, respectively, while
the dashdot line corresponds to $\alpha_s = 0$ and $B = 200$ MeV/fm$^3$.
Comparison between the dotdash line and those corresponding to $\alpha_s \neq 0$ 
suggests that, contrary to what stated by many authors (see, e.g., Steiner 
et al \cite{steiner}), perturbative gluon exchange, whose inclusion produces a 
sizable change of slope, cannot be simulated by adjusting
the value of the bag constant and must be explicitely taken into account.

\section{From nuclear matter to quark matter}
\label{phase_transition}

Early works on the possible occurrence of quark matter in neutron 
stars (e.g. Baym and Chin \cite{baym}) were based on the assumption that nuclear 
and quark matter were both charge neutral. As a consequence, the transition was 
described using Maxwell construction (see, e.g., Huang \cite{huang}) and
the resulting picture of the star consisted of a quark matter core surrounded 
by a mantle of nuclear matter, the two phases being separated by a sharp interface.

In the 90s, Glendenning (\cite{glend1,glend2}) pointed out that this assumption is 
too restrictive. More generally, the transition can proceed through the formation 
of a mixed phase of charged nuclear and quark matter, global neutrality being 
guaranteed by a uniform lepton background.

Equilibrium between charged phases of nuclear
matter ($NM$) and quark matter ($QM$) at $T=0$ requires the fulfillment of 
Gibbs conditions (see, e.g., Huang \cite{huang})
\begin{equation}
P_{NM}(\mu_{NM}^B,\mu_{NM}^Q) = P_{QM}(\mu_{QM}^B,\mu_{QM}^Q) \ ,
\label{eq:P}
\end{equation}
\begin{equation}
\mu_{NM}^B=\mu_{QM}^B \ \ \ \ , \ \ \ \ \mu_{NM}^Q=\mu_{QM}^Q \ ,
\label{eq:mu}
\end{equation}
where $P$ denotes the pressure, while $\mu^B$ and $\mu^Q$ are the chemical 
potentials associated with the two conserved quantities, namely 
baryonic number $B$ and electric charge $Q$. 

The above equations imply that, for any pressure ${\bar P}$, 
the projection of the surfaces
$P_{NM}(\mu^B,\mu^Q)$ and $P_{QM}(\mu^B,\mu^Q)$ onto the $P={\bar P}$ plane
defines two curves, whose intersection corresponds to the equilibrium 
values of the chemical potentials.
As the chemical potentials determine the charge densities of the two phases, 
the volume fraction occupied by quark matter, $\chi$, can then be obtained 
exploiting the requirement of {\it global} neutrality
\begin{equation}
\chi Q_{QM} + (1-\chi) Q_{NM} + \sum_\ell Q_\ell = 0 \ ,
\label{chargeneut}
\end{equation}
where $Q_{QM}$, $Q_{NM}$ and ${Q_\ell}$ denote the electric charge 
carried by nuclear matter, quark matter and leptons, respectively. 
From Eq.(\ref{chargeneut}) it follows that 
\begin{equation}
\chi = \frac{Q_{NM} + \sum_\ell Q_\ell}{Q_{NM} - Q_{QM}} \ ,
\label{def:chi}
\end{equation}
with $0 \leq \chi \leq 1$. Finally, the total energy density $\epsilon$ 
can be calculated using
 \begin{equation}
\epsilon = \chi \epsilon_{QM} + (1-\chi) \epsilon_{NM} \ ,
\label{endens0}
\end{equation}
and the EOS of state of the mixed phase can be cast in the standard 
form $P=P(\epsilon)$.

Requiring that the two phases be individually neutral, as in the pioneering work 
of Baym \& Chin (\cite{baym}), reduces the number
of chemical potentials to one, thus leading to the equilibrium conditions 
\begin{equation}
P_{NM}(\mu_{NM}^B) = P_{QM}(\mu_{QM}^B) \ ,
\label{eqm:P}
\end{equation}
\begin{equation}
\mu_{NM}^B=\mu_{QM}^B \ .
\label{eqm:mu}
\end{equation}
Within this scenario, charge neutral 
nuclear matter at baryon number density $n_B^{NM}$ coexists with charge neutral quark
matter at density $n_B^{QM}$, $n_B^{NM}$ and $n_B^{QM}$ being determined
by the requirements
\begin{equation}
\mu_B = \left( \frac{\partial \epsilon_{NM}}
{\partial n_B} \right)_{n_B=n^{NM}_B} =
\left( \frac{\partial \epsilon_{QM}}
{\partial n_B} \right)_{n_B=n^{QM}_B}\ .
\end{equation}
At $n^{NM}_B < n_B < n^{QM}_B$ pressure and chemical potential 
remain constant, the energy density is given by 
\begin{equation}
\epsilon = \mu_B n_B - P \ ,
\label{endens2}
\end{equation}
and the volume fraction occupied by quark matter grows linearly
with density according to
\begin{equation}
\chi = \frac{ \mu_B n_B - P - \epsilon_{NM}(n_{NM}^B) }
{\epsilon_{QM}(n^{QM}_B) - \epsilon_{NM}(n^{NM}_B)}. 
\end{equation}
Note that the above equation obviously implies that $0 \leq \chi \leq 1$, with
$\chi(n^{NM}_B)=0$ and $\chi(n^{QM}_B)=1$. 

\begin{figure}[hbt]
\centering
\includegraphics[width=7.0cm]{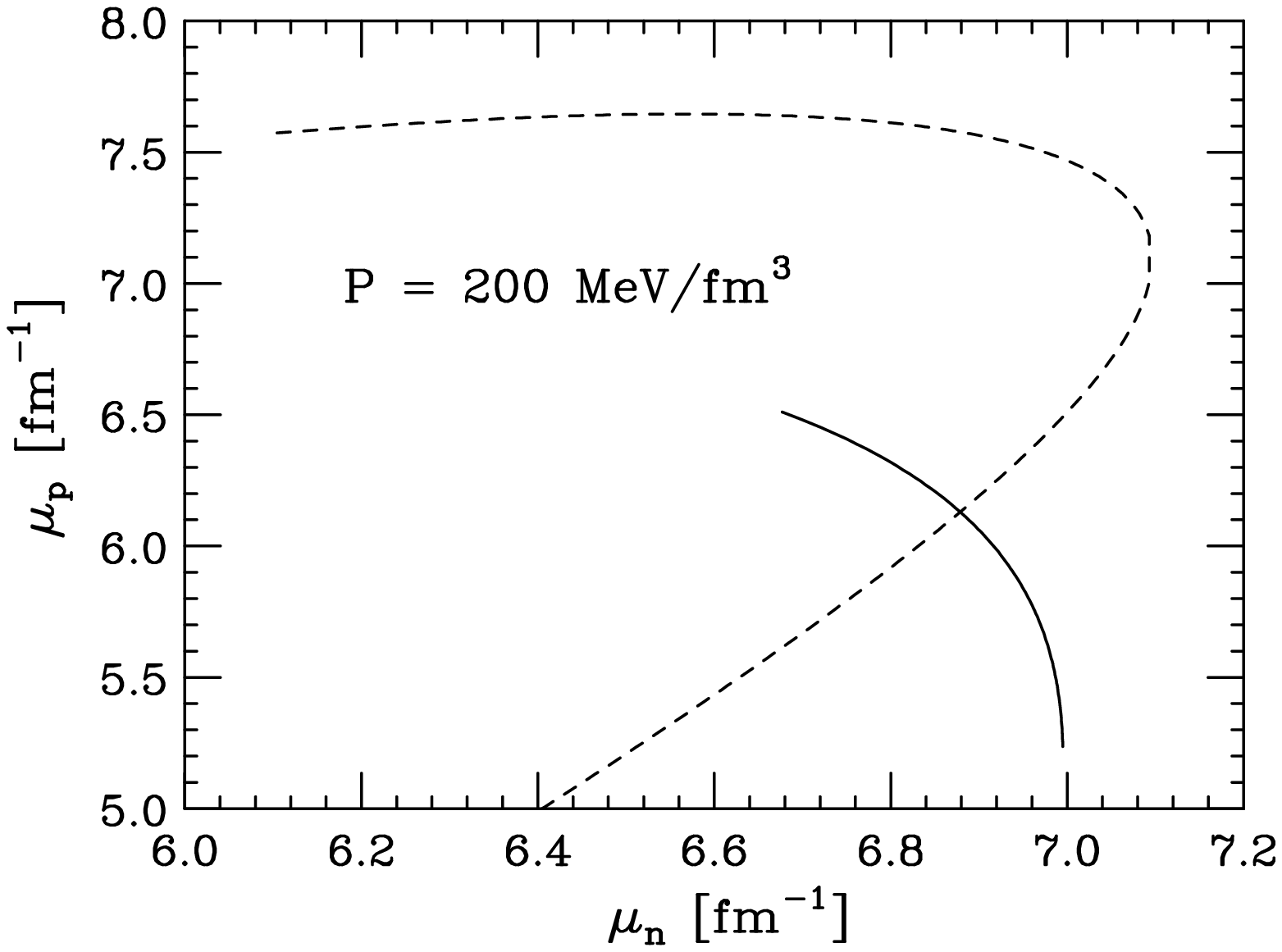}
\includegraphics[width=7.0cm]{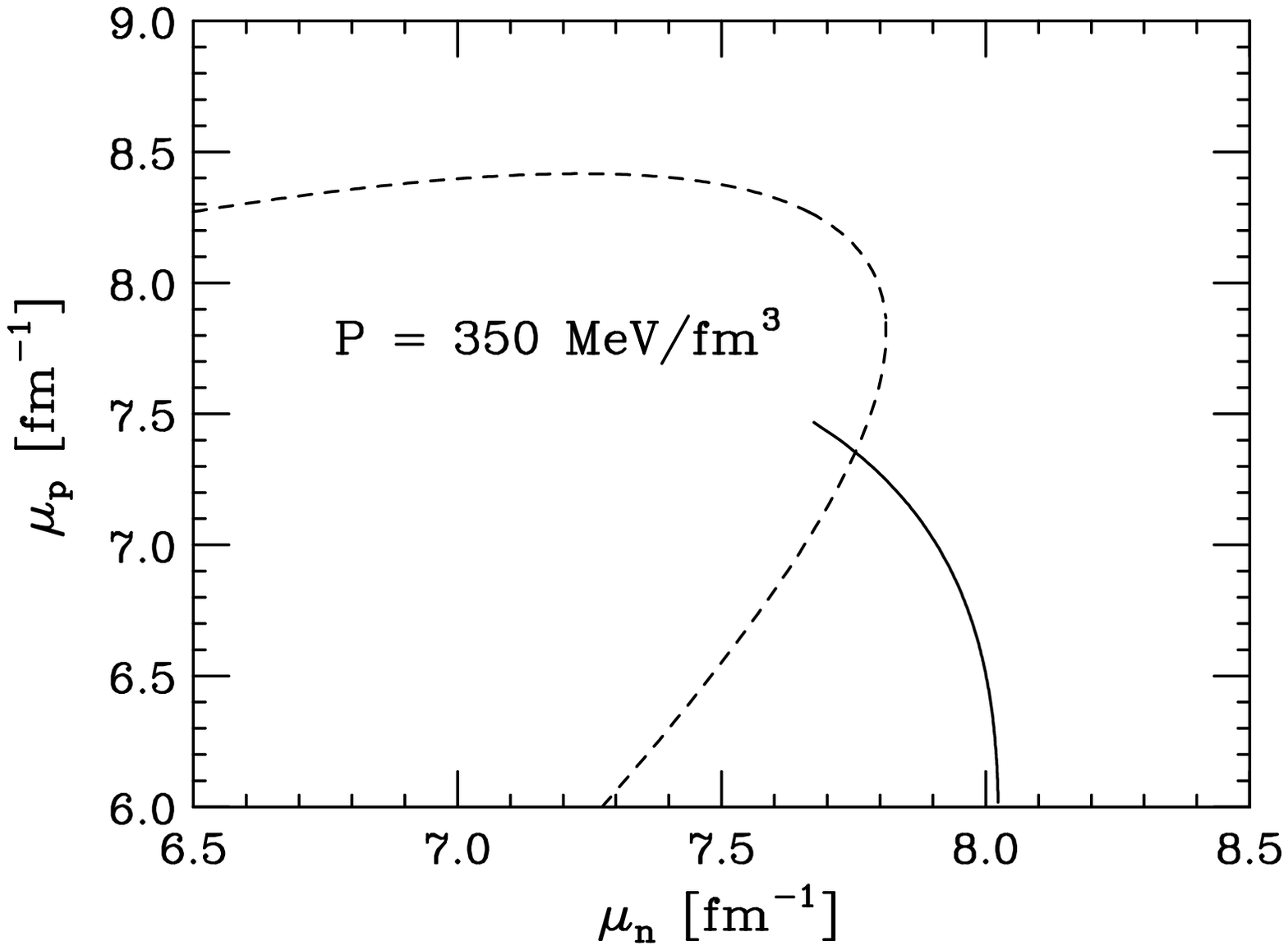}
\caption{Isobars $P(\mu_n,\mu_p)=200$ MeV/fm$^3$ (upper panel) and
$350$ MeV/fm$^3$ (lower panel) obtained using the APR EOS of nuclear
matter (solid lines) and the MIT bag model of quark matter, with
$\alpha_s = 0.5$ and $B = 200$ MeV/fm$^3$ (dashed lines).
The intersections determine the values of the chemical potentials
corresponding to equilibrium of the two phases according to Gibbs rules.
}
\label{findzeros}
\end{figure}

In the present work the intersection between the surfaces describing the 
pressure of nuclear 
and quark matter has been determined numerically, choosing as independent
variables, instead of $\mu^B$ and $\mu^Q$, the proton and neutron chemical
potentials $\mu_p$ and $\mu_n$. In nuclear matter they
are simply related to the lepton chemical potential through  the 
$\beta$-stability condition
$\mu_n - \mu_p = \mu_\ell$. In quark matter the chemical potentials 
of up and down quarks can be obtained from $\mu_p$ and $\mu_n$ inverting the 
relations
\begin{equation}
\mu_p = 2\mu_u + \mu_d
\label{def:mup}
\end{equation}
\begin{equation}
\mu_n = 2\mu_d + \mu_u \ ,
\label{def:mun}
\end{equation}
while the strange quark and lepton chemical potentials are dictated by the
conditions of weak equilibrium 
\begin{equation}
\mu_s = \mu_d \ ,
\end{equation}
\begin{equation}
\mu_d - \mu_u = \mu_\ell \ . 
\end{equation}

Fig. \ref{findzeros} illustrates the construction we have employed to determine
the values of $\mu_p$ and $\mu_n$ corresponding to equiliblrium between 
nuclear matter, described by the APR EOS, and quark matter, described by 
by the MIT bag model EOS with $\alpha_s = 0.5$ and $B = 200$ MeV/fm$^3$.
The region $P_{min} < P < P_{max}$ in which the isobars of nuclear and 
quark matter intersect defines the range of densities $n_{min} < n_B < n_{max}$ 
in which the mixed phase is energetically favored. At $n_B < n_{min}$ and 
$n_B > n_{max}$ the ground state consists of pure nuclear and quark matter, 
respectively.

\begin{figure}[hbt]
\centering
\includegraphics[width=7.0cm]{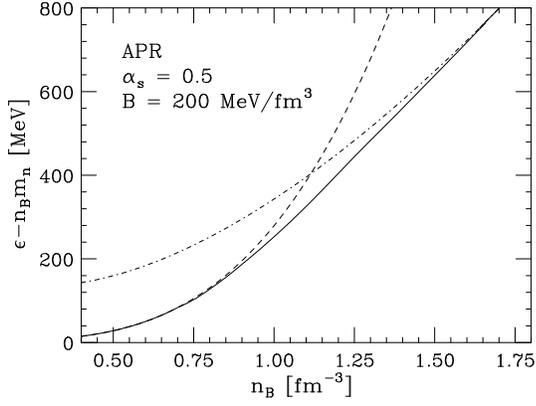}
\caption{
Dashed and dotdash lines show the energy density of charge neutral nuclear and 
quark matter in weak equilibrium, respectively. The bag model parameters have 
been set to B = 200 MeV/fm$^3$ and $\alpha_s = 0.5$. The solid line corresponds
to the mixed phase, obtained from Gibbs equilibrium conditions.
}
\label{phasetr:1}
\end{figure}

The phase transition between nuclear and quark matter, obtained setting 
B = 200 MeV/fm$^3$ and $\alpha_s = 0.5$, is illustrated in 
Fig. \ref{phasetr:1}. Dashed and dotdash lines show the dependence 
upon $n_B$ of the energy density of charge neutral nuclear and quark matter in 
weak equilibrium, respectively, while the solid line corresponds to the mixed 
phase. The latter turns out to be the ground state of neutron star matter
at densities $.7 \lsim n_B \lsim 1.7$ fm$^{-3}$. 

\begin{figure}[hbt]
\centering
\includegraphics[width=7.0cm]{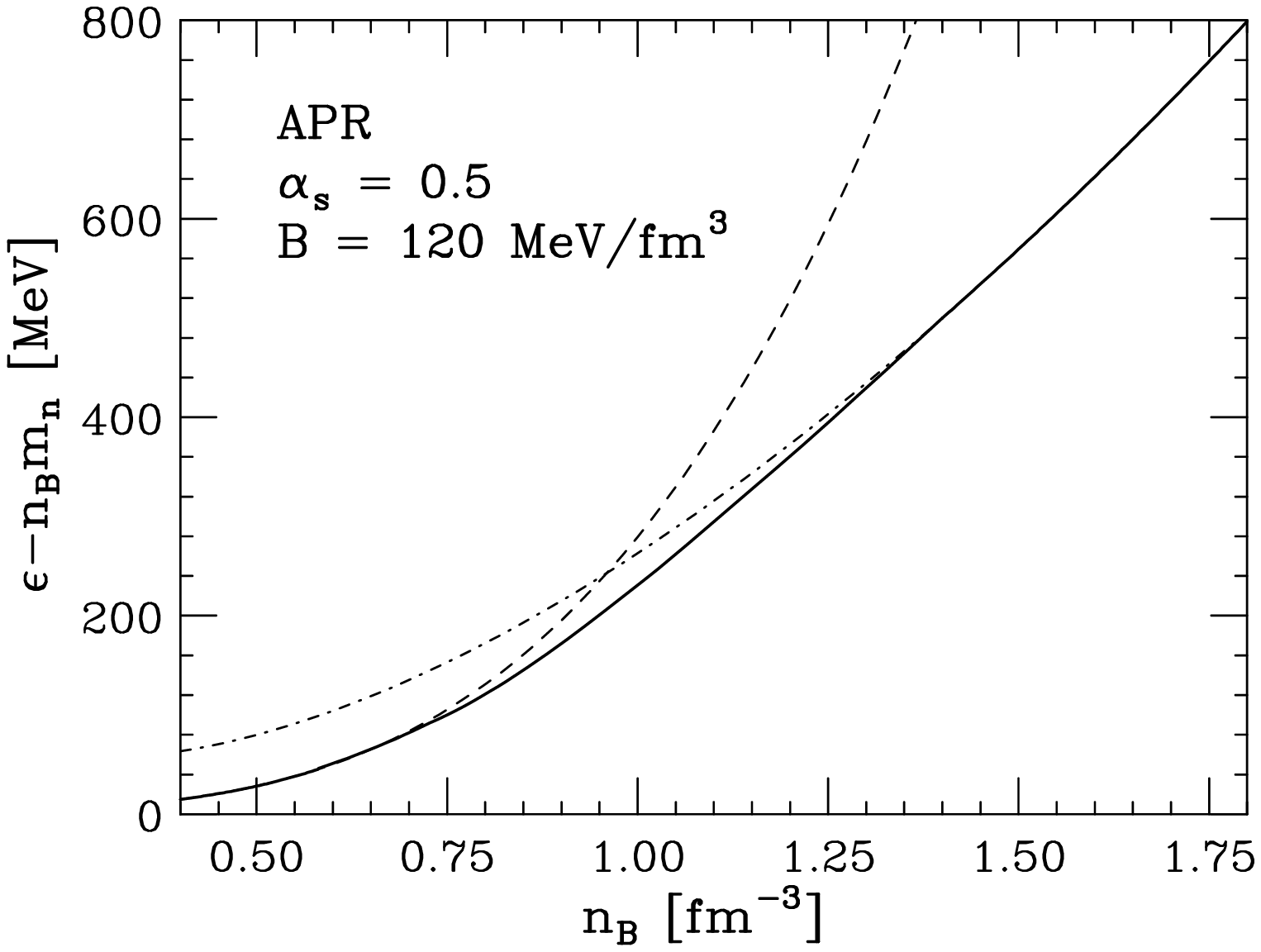}
\includegraphics[width=7.0cm]{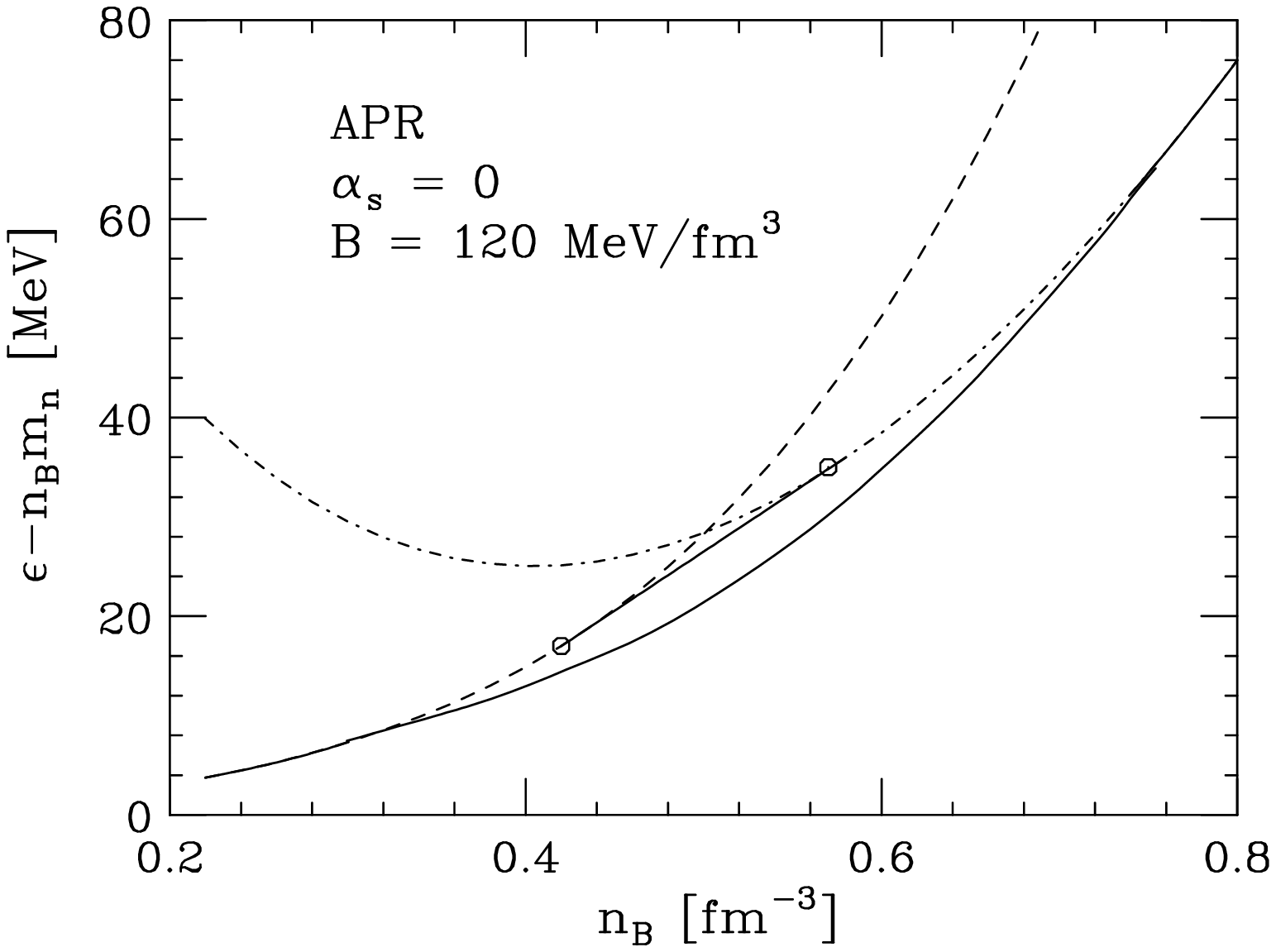}
\caption{
As in Fig. \protect\ref{phasetr:1}, but with the EOS of quark matter obtained using
different values of the MIT bag model parameters. Upper panel:  B = 120 MeV/fm$^3$ 
and $\alpha_s = 0.5$; lower panel  B = 120 MeV/fm$^3$ and $\alpha_s = 0.$.
The straight line in the lower panel is the double tangent resulting 
from Maxwell cosntruction, while the open circles show the extrema of the 
coexistence region. 
}
\label{phasetr:2}
\end{figure}

The dependence of our results upon the MIT bag model parameters can be gauged from
the upper panel of 
Fig. \ref{phasetr:2}. It clearly appears that a lower value of the bag constant, 
corresponding to a softer quark matter EOS, leads to the appearance of the mixed phase 
at lower density. Keeping $\alpha_s = 0.5$ and setting B = 120 MeV/fm$^3$ 
one finds that the mixed phase is energetically favored in the 
range $.6 \lsim n_B \lsim 1.4$ fm$^-3$.
An even larger effect, illustrated by the lower panel of Fig. \ref{phasetr:2} is 
obtained with B = 120 MeV/fm$^3$ and $\alpha_s = 0$, 
 i.e. neglecting perturbative gluon exchange altogether. 
For this case we also show the results obtained from Maxwell
construction, leading to the cohexistence of charge neutral nuclear matter at 
$n_B = .42$ fm$^{-3}$ and charge neutral quark matter at $n_B = .57$ fm${^-3}$.
This cohexistence region is to be compared to the region of stability of the 
mixed phase, corresponding to $.22 \lsim n_B \lsim .75$ fm$^{-3}$.

\section{Structure and stability of the mixed phase}
\label{surface}

The results of the previous Section suggest that, irrespective of the
details of the EOS, the transition from nuclear to quark matter proceed through the 
formation of a mixed phase. 
However, two issues relevant to both the appearance and 
the stability of the mixed phase, not taken into account in the discussion of
Section \ref{phase_transition}, need to be further analyzed. 

Consider a mixed phase consisting of droplets of quark matter immersed in
$\beta$-stable nuclear matter, global charge neutrality being guaranteed 
by a lepton background. This picture is obvioulsy based on the assumption
that the appearance of the charged droplets do not significantly affect the 
space distribution of the leptons, i.e. that the Debye screening length 
$\lambda_D$ be large compared to both the typical size of the droplets and their 
separation distance (Heiselberg et al \cite{HPS93}, Heiselberg 
\& Hjort-Jensen \cite{HMJ2000}). If this condition is not satisfied the
lepton background is distorted in such a way as to screen electrostastic
interactions. 

The estimates of $\lambda_D$ reported by
Heiselberg et al (\cite{HPS93}) suggest that screening effects 
can be disregarded if the structures appearing in the 
mixed phase of quark and nuclear matter have typical size and 
separation distance of $\sim 10$ fm. The results of our calculations, 
that will be discussed discussed later in this Section, show that this 
appears indeed to be the case.

The second issue deserving consideration is the stability of the 
mixed phase, i.e. the question wheter or not its energy is lower than the energy 
of the cohexisting phases of nuclear and quark matter. 

Formation of a spherical droplet of quark matter requires the energy   
\begin{equation}
{\mathcal E}_D = {\mathcal E}_C+{\mathcal E}_S\ .
\label{efinite}
\end{equation}
where the surface contribution ${\mathcal E}_S$ 
is parametrized in terms of the surface tension $\sigma$ according to
\begin{equation}
{\mathcal E}_S =  \sigma \ 4 \pi R^2\ ,
\end{equation}
$R$ being the droplet radius. The electrostatic energy ${\mathcal E}_C$ can be cast 
in the form
(Ravenhall et al \cite{RPW}, Pethick et al \cite{PRL})
\begin{equation}
{\mathcal E}_C = \frac{3}{5} \frac{Q^2}{R} \left( 1
- \frac{3}{2} u^{1/3} + \frac{1}{2} u \right)\ ,
\end{equation}
with $u=(R/R_c)^3$, $R_c$ being the radius of the Wigner-Seitz cell. Note that the 
first term in the right hand side of the above equation is the self energy of a 
droplet of radius $R$ and charge $Q$ obtained from Gauss law. The electric charge
$Q$ is given by
\begin{equation}
Q = \frac{4 \pi R^3}{3}\ (\rho_{QM} - \rho_{NM}) \ ,
\end{equation}
$\rho_{QM}$ and $\rho_{NM}$ being the charge densities of quark matter and nuclear 
matter, respectively. Minimization of the energy density 
$\epsilon = 3{\mathcal E}_D/4\pi R_c^3$ 
with respect to the droplet radius yields
\begin{equation}
{\mathcal E}_S = 2 {\mathcal E}_C
\end{equation}
and
\begin{equation}
R = \left[ \frac{4 \pi (\rho_{QM} - \rho_{NM})^2}{3 \sigma}\ f_3(u) \right]^{-1/3} \ ,
\label{drop:radius}
\end{equation}
where 
\begin{equation}
f_3(u) = \frac{1}{5} \left( 2 - 3u^{1/3} + u \right)\ .
\end{equation}

As the density increases, the droplets begin to merge and give rise to structures of 
variable dimensionality, changing firts from spheres into rods ({\it spaghetti})
and eventually into slabs ({\it lasagna}). At larger densities the volume fraction 
occupied by quark matter exceeds $1/2$, and the role of the two phases is reversed. 
Nuclear matter, initially arranged in slabs, turns into rods and 
shperes that finally dissolve in uniform charge neutral quark matter.

The energy density needed for the formation of the structures appearing in the 
mixed phase has been obtained by Ravenhall et al (\cite{RPW}) in the 
context of a study of matter in the neutron star inner crust. It can be cast in 
the concise form
\begin{equation}
\epsilon_C + \epsilon_S = 6 \pi u \left[ \frac{\alpha}{4 \pi} \sigma^2 d^2 
\left( n_B^{NM} - n_B^{QM} \right)^2 f_d(u) \right]\ ,
\label{pasta1}
\end{equation}
where $\alpha$ is the fine structure constant, $u$ is the volume fraction occupied 
by the less abundant phase (i.e. $u = \chi$ for $\chi < 1/2$, $u=1-\chi$ for 
$\chi \geq 1/2$) and 
\begin{equation}
f_d(u) = \frac{1}{d+2} \left[ \frac{1}{d-2} \left( 2 - d u ^{1-2/d} \right)
 + u \right]\ .
\label{pasta2}
\end{equation}
For $d=1,2$ and $3$ Eqs.(\ref{pasta1}) and (\ref{pasta2}) provide the correct energy density for 
the case of slabs, rods and spheres, respectively.

For $\sigma = 0$ both surface and Coulomb energies vanish, and the energy density 
of the mixed phase is given by Eq.(\ref{endens0}), while for $\sigma \neq 0$ 
\begin{equation}
\epsilon(\sigma) = \epsilon(\sigma=0) + \epsilon_C + \epsilon_S \ .
\label{endens1}
\end{equation}
The mixed phase is energetically favorable if $\epsilon(\sigma)$ is less than the 
energy density obtained from Maxwell construction, given by Eq.(\ref{endens2}).

The value of the surface tension at the interface between nuclear and quark matter is not
known. It has been estimated using the MIT bag model and ignoring gluon exchange 
(Berger \& Jaffe \cite{sigma1}, \cite{sigma2}). Assuming a strange 
quark has mass of $\sim 150 $ MeV Berger \& Jaffe predict $\sigma \sim 10$ MeV/fm$^2$.
To quantitatively investigate the stability of the mixed phase, we have calculated 
$\Delta_\epsilon = \epsilon(\sigma) - \epsilon(0)$ for different values of 
$\sigma$, ranging from 2 MeV/fm$^2$ to 10 MeV/fm$^2$.

\begin{figure}[hbt]
\centering
\includegraphics[width=7.0cm]{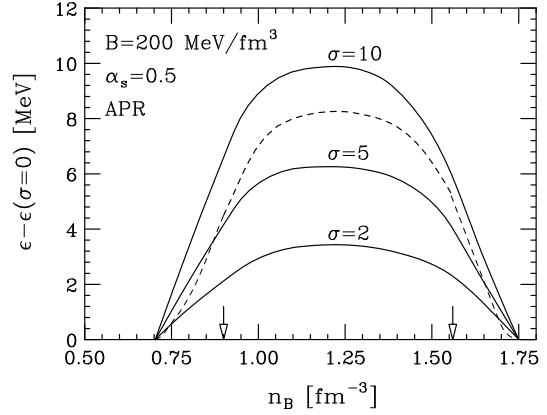}
\caption{
The solid lines correspond to the difference 
$\Delta_\epsilon = \epsilon(\sigma) - \epsilon(0)$ (see Eq.(\ref{endens1})), 
evaluated for $\sigma = 10$, $5$ and $2$ MeV/fm$^3$. The dashed line
shows the difference $\widetilde{\Delta}$ between the energy density resulting from 
Maxwell conatruction and $\epsilon(0)$. The arrows mark the limits of
the cohexistence region. 
Nuclear and quark matter are described by the APR EOS  
and the MIT bag model EOS, with $\alpha_s = 0.5$
and $B = 200$ MeV/fm$^3$, respectively. 
}
\label{surf:1}
\end{figure}

For any given value of the baryon number density $n_B$, 
the energy density of Eqs.(\ref{pasta1})-(\ref{pasta2}) has been calculated using the nuclear 
and quark matter densities determined according to the procedure described in Section 
\ref{phase_transition} and carrying out a numerical 
minimization with respect to the value of the dimensionality parameter $d$. 
As $n_B$ increases, the resulting values of $d$ change initially from $\sim 3$ to $\sim 2$ 
and $\sim 1$ and then again to $\sim 2$ and finally to $\sim 3$. 
For example, in the case illustrated by Fig. \ref{surf:1}, and corresponding to
$\sigma = 10$ MeV/fm$^2$, we find that sherical droplets of quark matter ($d \sim 3$) appear at 
$n_B \sim .75$ fm$^{-3}$ and turn into rods ($d \sim 2$) and slabs ($d \sim 1$) at
$n_B \sim .95$ and $ \sim 1.2$ fm$^{-3}$, respectively. For larger densities quark matter 
becomes 
the dominant phase (i.e. $\chi > 1/2$): at $n_B \sim 1.5$ and $\sim 1.7$ fm$^{-3}$ the mixed
phase features rods ($d \sim 2$) and droplets ($d \sim 3$) of nuclear matter that 
eventually dissolve in the quark matter background.

\begin{figure}[hbt]
\centering
\includegraphics[width=7.0cm]{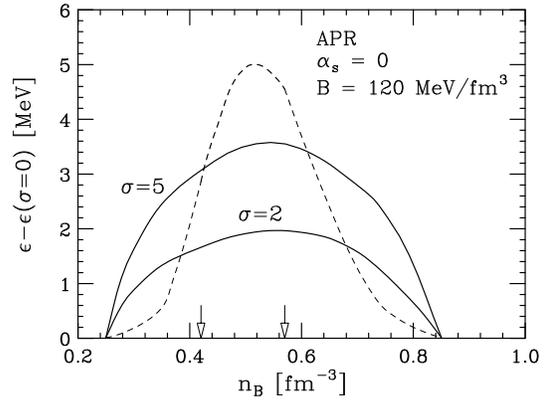}
\caption{
Same as in Fig. \protect\ref{surf:1}, but with $\alpha_s = 0$ and 
$B = 120$ MeV/fm$^3$. The solid lines show to the difference
$\Delta_\epsilon = \epsilon(\sigma) - \epsilon(0)$ (see Eq.(\ref{endens1})),
evaluated for $\sigma = 5$ and $2$ MeV/fm$^3$, respectively.
}
\label{surf:2}
\end{figure}

Our results are summarized in Figs. \ref{surf:1} and \ref{surf:2},
that correspond to different choices of the MIT bag model parameters.
The solid lines show the $n_B$ dependence of the calculated $\Delta_\epsilon$ 
for different values of the surface tension $\sigma$. The dashed line represents
the difference $\widetilde{\Delta}_\epsilon = \epsilon_M - \epsilon(0)$, where
$\epsilon_M$ is the energy density obtained from Maxwell construction. 
For any given value of the surface tension, the mixed pase is favorable if the 
corresponding solid line lies below the dashed line.

The results of Fig. \ref{surf:1}, corresponding to $B=200$ MeV/fm$^3$ and
$\alpha_s=0.5$, show that the mixed phase, while being always the lowest energy phase
for $\sigma=2$ MeV/fm$^2$, becomes to be energetically unfavorable at some densities for 
$\sigma \gsim 5$ MeV/fm$^2$. For $\sigma=10$ MeV/fm$^2$ cohexistence of charge 
neutral phases of nuclear and quark matter turn out to favorable over the whole
density range.

To gauge the dependence upon the MIT bag model parameters we have
repeated the calculations setting $B=120$ MeV/fm$^3$ and $\alpha_s=0$. 
The results of Fig. \ref{surf:2} show that for $\sigma$ in the range 
$2-5$ MeV/fm$^2$ the mixed phase is energetically favorable over a density region
larger than the cohexistence region predicted by Maxwell construction. 

Finally, we return to the problem of the comparison between the Debye 
screening length and the typical size of the structures appearing
in the mixed phase. Our results suggest that the condition outlined in the 
work of Heiselberg et al (\cite{HPS93}) are indeed fulfilled.
For example, in the case $B=200$ MeV/fm$^3$ and
$\alpha_s=0.5$ we find that in the region of $\chi \ll 1$, corresponding to
formation of droplets of quark matter, the droplets radius given by 
Eq.(\ref{drop:radius}) is $\sim 2-3$ fm.

\section{Implications for star structure}
\label{star}

Plugging the EOS $P=P(\epsilon)$ into the Tolman Oppenheimer 
Volkoff (TOV) equations (Tolman \cite{T}, Oppenheimer \& Volkoff \cite{OV})
\begin{equation}
\frac{dP(r)}{dr} = - G\
\frac{ \left[ \epsilon(r) + P(r) \right]
 \left[ M(r) + 4 \pi r^2 P(r) \right] }
{ r^2 \left[ 1 - 2 G M(r) /r \right] }\ ,
\label{TOV1}
\end{equation}
where G denotes the gravitational constant, and
\begin{equation}
M(r) = 4 \pi \int_0^r {r^\prime}^2 d{r^\prime}
\epsilon({r^\prime})\  , 
\label{TOV2}
\end{equation}
one can obtain the properties of the stable configurations of a nonrotating neutron star. 
Eqs.(\ref{TOV1}) and (\ref{TOV2}) are solved by integrating outwards with the initial 
condition $\epsilon(r=0) = \epsilon_c$. For any given value of the central 
density, $\epsilon_c$, the star radius $R$ is determined by the condition $P(R)=0$ and
its mass $M=M(R)$ is given by Eq.(\ref{TOV2}).

\begin{figure}[hbt]
\centering
\includegraphics[width=7.0cm]{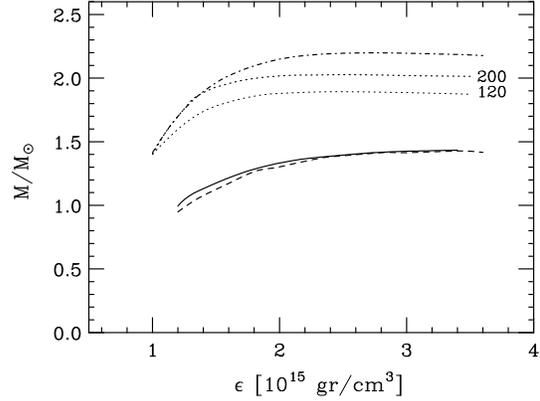}
\caption{
Relation between neutron star mass and central energy density for different EOS.  
Dotdash line: pure nuclear matter (APR EOS); dotted lines: nuclear matter (APR EOS) 
and quark matter (MIT bag model with $B=$ 120 and 200 MeV/fm$^3$ and $\alpha_s= 0.5$). 
The phase transition is described according to Gibbs rules;
 solid line: same as the dotted line, but with $B= 120$ MeV/fm$^3$ 
and $\alpha_s= 0.5$;
 dashed line: same as the solid line, but with the phase transition described 
using Maxwell construction.
}
\label{star:1}
\end{figure}

The occurrence of the transition to quark matter makes the EOS softer, thus 
leading to a lower value of the maximum mass. 
In Fig. \ref{star:1} we compare the mass-central energy density relations 
obtained using the APR EOS only to that obtained allowing for a 
transition to quark matter with $\alpha_s= 0.5$ and $B=$ 120 and 200 MeV/fm$^3$. 
The transition is described according to Gibbs conditions, neglecting
surface and Coulomb effects. We find $M_{max} =$ 2.20 $M_\odot$ for 
the star made of nuclear matter oly and $M_{max} =$ 1.89 and 2.03 $M_\odot$ 
for the hybrid stars corresponding to $B=$ 120 and 200 MeV/fm$^3$, respectively.

In Fig. \ref{star:1} we also compare the $M(\epsilon_c)$ curves obtained 
setting $B= 120$ MeV/fm$^3$ and $\alpha_s= 0$ and adopting
either Gibbs or Maxwell picture.
Whether the phase transition proceeds through the appearance of a mixed phase
or charge neutral cohexisting phases does not appear to significanltly affect 
the mass-central energy density relation. On the other hand, neglecting
perturbative gluon exchange results in a rather low maximum mass, 
$M_{max} \sim$ 1.4 $M_\odot$, barely compatible with the measured 
neutron star masses. 

\begin{figure}[hbt]
\centering
\includegraphics[width=7.0cm]{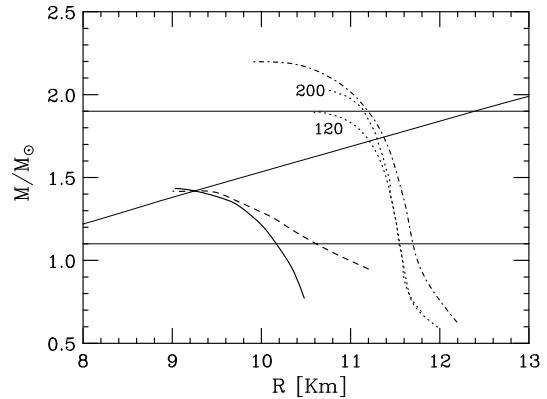}
\caption{
Mass radius relation for different EOS. The meaning of the curves is the same as
in Fig. \protect\ref{star:1}. The horizontal lines correspond to the observational 
limits on neutron star mass, whereas the third straight line is the mass-radius 
relation resulting from the gravitational redshift measurement of 
Cottam et al (\protect\cite{redshift}).
}
\label{star:2}
\end{figure}

The neutron star mass-radius relations associated with
the $M(\epsilon_c)$ curves of Fig. \ref{star:1}, displayed in Fig. \ref{star:2}, 
show that in this case using Maxwell construction instead of Gibbs rules 
produces a visible effect. It has to be noted that all EOS predict 
the existence of stable star configurations with masses
in the range allowed by observation (Thorsett \& Chakrabarty 
\cite{nsm1}, Qaintrell et al \cite{nsm2}), as well as a $M(R)$ relation 
compatible with that resulting from the gravitational red shift measurement 
of Cottam et al (\cite{redshift}).

\begin{figure}[hbt]
\centering
\includegraphics[width=7.0cm]{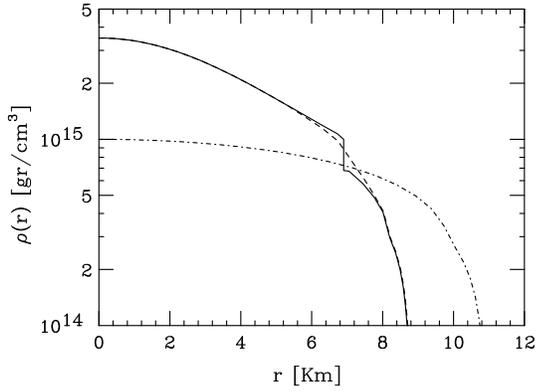}
\caption{
Density profiles of a neutron star of mass $\sim$ 1.4 $M_\odot$ resulting 
from different models. Dashed line: APR EOS and 
MIT bag model EOS with $\alpha_s=0$ and $B =$ 120 MeV/fm$^3$. Phase transition 
described allowing for the presence of the mixed phase. Solid line: same 
as the solid line, but with the phase transition described according to 
Maxwell construction. Dotdash line: pure nuclear matter with APR EOS.
}
\label{star:3}
\end{figure}

Fig \ref{star:3} shows that 
different descriptions of the phase transition lead to remarkably different
 star density profile. While in presence of the mixed phase the density 
is a smooth function of the distance from the star center, Maxwell construction 
leads to the appearance of a disontinuity. For comparison, we also show the 
profile of a star of the same mass, $\sim$ 1.4 $M_\odot$, made of pure 
nuclear matter described by the APR EOS.

The discontinuous behavior can be easily understood noting that TOV equations 
(\ref{TOV1}) and (\ref{TOV2}) require that the pressure $P(r)$ be a monotonically 
decreasing function. It follows that if the pressure is the same for two different
values of density, as in the phase transition \`a la Maxwell, they must necessarily
correspond to the same value of $r$.

In the cohexisting phases scenario, the transition only takes place in stars whose
central density exceeds the density of the quark matter phase. These star configurations
turn out to be marginally stable, their mass being close to the maximum mass.
For example, setting $B=$ 200 MeV/fm$^3$ and $\alpha_s=0.5$ we find that the 
transition only occurs in stars having mass $\sim 2.0 M_\odot$. The radius of 
the quark matter core is pretty small ($\sim$ 1 Km), while the density 
jump is large, going from $1.8\times10^{15}$ g/cm$^3$ to $3.9\times10^{15}$ g/cm$^3$.
These results are to be compared to those obtained in the mixed phase scenario 
when Coulomb and surface effects are neglected.
In this case there is no jump and the density varies smoothly. At the center
of a star of mass $\sim 2.0 M_\odot$, corresponding to energy-density 
$\sim 2.4\times10^{15}$ g/cm$^3$ the volume fraction occupied by quark matter 
reaches $\chi \sim 30$ \%.

\section{Conclusions}
\label{conclusions}

We have carried out a study of the transition from nuclear matter to 
quark matter in the inner core of neutron stars, aimed at assessing 
whether the appearance of a mixed phase is energetically favorable and
how the emerging picture depends upon the parameters entering the MIT bag 
model EOS. 

In order to minimize the uncertainty associated with the description of the 
nuclear matter phase, we have adopted a EOS obtained from an {\it ab initio} 
calculation, based on a dynamical model stongly contrained by experimental 
data and not involving any adjustable parameters (Akmal et al \cite{apr2}).

Our results show that the effect of perturbative gluon exchange on the MIT 
bag model 
EOS is large and cannot be accounted for adjusting the value of the bag constant $B$.
Neglecting interactions between quarks leads to a considerable softening of the 
EOS, resulting in a drastically lower transition density.

The fact that the setting $\alpha_s=0$ lead to a much softer EOS is reflected
by the rather small value of the maximum neutron star mass, barely exceeding
the canonical value of 1.4 $M_\odot$. 

The stability of the mixed phase turns out to be strongly affected by surface and 
Coulomb effects. Using the softer quark matter EOS 
($\alpha_s=0, B=120$ MeV/fm$^3$) we find that even a very small 
value 
of the surface tension, $\sigma = 2$ MeV/fm$^3$, produces a narrowing of the 
density region spanned by the mixed phase. With the harder EOS 
($\alpha_s=0.5, B=200$ MeV/fm$^3$) the cohexistence
of neutral phases of nuclear and quark matter is energetically favored at all
densities for $\sigma = 10$ MeV/fm$^3$.

Comparison with the results of Heiselberg et al suggests that surface and Coulomb 
effect become larger when a realistic EOS is employed to describe the nuclear matter
phase. 

While the mass-central density relation appears to be largely unaffected by the 
occurrence of the mixed phase, its effect can be clearly seen in the $M(R)$ 
curve. However, as measurements of neutron star radii are plagued
by large uncertainties, this feature is not likely to be exploitable to 
extract clearcut information from observations.

The most striking difference between the Maxwell and Gibbs picture of
the phase transition shows up in the neutron star density profile, 
which in the case of transition at constant pressure exhibits
a sharp discontinuity. 

The presence of a density jump is known to 
affect neutron star dynamics, leading to appearance of a class of 
nonradial oscillation modes, called $g$-modes, associated 
with emission of gravitational radiation. 

Early investigations of the $g$-modes 
focused on the discontinuities produced by the changes of
chemical composition in the low density region of the neutron star crust, 
corresponding to a fractional distance from the surface $\lsim$~10~\%
(Finn \cite{finn}, McDermott \cite{mcdermott}). 
These studies have been recently extended to the case of 
$g$-modes produced by a discontinuity located 
at much larger density and involving a much larger density jump, 
such as those associated with the transition to quark matter  
 (Miniutti et al \cite{gmodes}). Based on the results of calculations 
carried out using siple polytropic EOS, Miniutti et al (\cite{gmodes}) argue that 
 a simultanuoeus measurements of the
frequencies of the fundamental $f$-mode and the $g$-mode would provide 
information on both size and location of the discontinuity.

Although it is unlikely that the
first generation of laser interferometric antennas will detect gravitational waves 
emitted by an oscillating  neutron star,
 the new detectors currenltly under investigation 
(see, e.g., the EURO proposal (\cite{EURO})) are 
expected to be much 
more sensitive at the relevant frequencies above 1-2 kHz.
Hopefully, information on the neutron star matter EOS and the nature of 
the transition to quark matter may come gravitational wave astronomy.

\begin{acknowledgements}
The authors are deeply indebted to Ignazio Bombaci, Adelchi Fabrocini, 
Stefano Fantoni, Valeria Ferrari and Vijay Pandharipande for a number of 
usefuls discussions on issues related to the subjet of this paper. 
\end{acknowledgements}

\end{document}